\documentclass[referee,sn-vancouver-num]{sn-jnl}

\usepackage{graphicx}%
\usepackage{longtable}
\usepackage{array}%
\usepackage{multirow}%
\usepackage{amsmath,amssymb,amsfonts}%
\usepackage{amsthm}%
\usepackage{mathrsfs}%
\usepackage[title]{appendix}%
\usepackage{xcolor}%
\usepackage{textcomp}%
\usepackage{manyfoot}%
\usepackage{booktabs}%
\usepackage{algorithm}%
\usepackage{algorithmicx}%
\usepackage{algpseudocode}%
\usepackage{listings}%
\usepackage{tikz}%
\usetikzlibrary{shapes.geometric, arrows.meta, positioning}%

\theoremstyle{thmstyleone}%
\theoremstyle{thmstyletwo}%

\theoremstyle{thmstylethree}%

\begin{document}

\title[Article Title]{Bridging openEHR and OMOP: Expanded Mappings and Systematic Analysis of Semantic and Structural Limitations in the OMOP CDM}

\author*[1,2,3,4]{\fnm{Severin} \sur{Kohler}}\email{severin.kohler@bih-charite.de}

\author[5]{\fnm{Diego} \sur{Boscá}}

\author[6,7]{\fnm{Falk} \sur{Meyer-Eschenbach}} 

\author[8]{\fnm{Prabash} \sur{Galgane Banduge}}

\author[9,10]{\fnm{Somayeh} \sur{Abedian}}

\author[4]{\fnm{Michael} \sur{Marschollek}}

\author[1,3,11]{\fnm{Roland} \sur{Eils}} 

\affil[1]{\orgdiv{Digital Health Center}, \orgname{Berlin Institute of Health at Charité - Universitätsmedizin Berlin}, \orgaddress{\city{Berlin}, \country{Germany}}}

\affil[2]{\orgdiv{Institute for Mathematics and Informatics}, \orgname{Freie Universität Berlin}, \orgaddress{\city{Berlin}, \country{Germany}}}

\affil[3]{\orgdiv{Health Data Science Unit}, \orgname{Heidelberg University Hospital and BioQuant}, \orgaddress{\city{Heidelberg}, \country{Germany}}}

\affil[4]{\orgname{Peter L. Reichertz Institute for Medical Informatics of TU Braunschweig and Hannover Medical School}, \orgaddress{\city{Hannover}, \country{Germany}}}

\affil[5]{\orgname{VeraTech for Health}, \orgaddress{\city{Valencia}, \country{Spain}}}

\affil[6]{\orgname{Institute of Medical Informatics, Charité - Universitätsmedizin Berlin}, \orgaddress{\city{Berlin}, \country{Germany}}}

\affil[7]{\orgdiv{Clinical Study Center}, \orgname{Berlin Institute of Health at Charité - Universitätsmedizin Berlin}, \orgaddress{\city{Berlin}, \country{Germany}}}

\affil[8]{\orgdiv{Maastricht DataHub}, \orgname{Maastricht University}, \orgaddress{\city{Maastricht}, \country{Netherlands}}}

\affil[9]{\orgdiv{Ludwig Boltzmann Institute for Digital Health and Prevention}, \orgaddress{\city{Salzburg}, \country{Austria}}}

\affil[10]{\orgname{Institute of Health Policy, Management and Evaluation, Dalla Lana School of Public Health, University of Toronto}, \orgaddress{\city{Toronto}, \country{Canada}}}

\affil[11]{\orgname{Intelligent Medicine Institute}, \orgdiv{Fudan University}, \orgaddress{\city{Shanghai}, \country{China}}}

\abstract{
\textbf{Background:}
Interoperability between clinical and research systems is essential for secondary use of EHR data. The openEHR standard provides structured, model-driven clinical information, while the OMOP Common Data Model (CDM) supports large-scale observational analytics. The Eos engine and OMOP Conversion Language (OMOCL) previously introduced a standards-based transformation approach, but limited support for internal value sets, rigid visit generation logic, and incomplete mapping coverage restricted applicability.

\textbf{Methods:}
A new generation of Eos and OMOCL was implemented to address these limitations. New functionality enables mapping of internal openEHR value sets via conceptMaps, supports visit occurrence generation using Archetype Query Language (AQL), and expands the international archetype mapping library. Mapping coverage, terminology completeness and domain distribution were evaluated. Limitations of the OMOP CDM were assessed over the 198 archetype mappings with a directed content analysis. A codebook from an initial pass identified vocabulary gaps, structural fragmentation and semantic field overloading as the measurable categories.

\textbf{Results:}
A total of 198 openEHR archetypes were mapped, covering all stable archetypes in the international Clinical Knowledge Manager with OMOP-equivalent tables. Overall, 8.59\% of primary concept identifiers could not be linked to OMOP standard terminologies, indicating persistent vocabulary gaps. Most mappings targeted the Measurement (49.8\%) and Observation (41.7\%) domains. The enhanced conceptMap logic reduced semantic information loss, and AQL-based visit generation improved adaptability across heterogeneous openEHR implementations. Structural analysis showed that comprehensive representation of coherent clinical concepts often required fragmentation across multiple loosely connected OMOP tables affecting 87.2\% of the openEHR nodes. Several OMOP fields are used to represent several distinct clinical qualifiers, necessitating heterogeneous extract-transform-load (ETL) decisions and reducing cross-site comparability.

\textbf{Conclusions:}
This new generation of the open-source Eos and OMOCL framework strengthens standards-based interoperability between openEHR and OMOP. However, the mappings reveal structural and semantic limitations within the OMOP CDM that may introduce fragmentation and ambiguity affecting downstream analytics. Greater convergence between openEHR's archetype-driven model and OMOP's analytics-oriented ecosystem may provide a more sustainable foundation for semantically robust secondary use of clinical data.
}



\maketitle

\section{Background}
Despite decades of digitization, routine reuse of Electronic Health Record (EHR) data for research is still limited by the lack of semantic and structural interoperability \cite{de2022semantic, palojoki_semantic_2024}.
As a result, EHR data are characterized by heterogeneous structures and inconsistent semantics \cite{Coorevits2013, Commission2013, Prokosch2009}, making large scale analytics dependent on complex and resource intensive integration pipelines \cite{Kimball2013}. This lack of uniform data models and standardized meaning remains one of the central barriers to both primary and secondary use of health data \cite{de2022semantic, Commission2013, dentler2013barriers}.

To ensure the comparability and analytical readiness of such heterogeneous data sources for secondary use, standardized data models are required. The Observational Medical Outcomes Partnership (OMOP) Common Data Model (CDM) \cite{hripcsak2015observational} is one of the most widely adopted standards for this purpose, providing a harmonized structure and terminology system to facilitate reproducible analyses across institutions and countries \cite{reinecke2021usage}.

Despite ongoing advances in interoperability standards such as openEHR \cite{kalra2005openehr} and Fast Healthcare Interoperability Resources (FHIR) \cite{bender2013hl7}, transforming clinical data into OMOP CDM remains a resource-intensive and technically demanding process \cite{quiroz2022extract}. Each institution typically needs to create and maintain its own Extract Transform Load (ETL) processes to bridge local data representations and the OMOP schema. This lack of reusability is a barrier to large-scale data integration and can lead to inconsistencies in how health data are  interpreted and transformed into OMOP. Reusable, standards-based transformation pipelines can help overcome these challenges by offering transparent, reproducible, and community-driven mappings.

Among existing interoperability standards, openEHR has been identified as particularly suitable for the structured, routine documentation of clinical and administrative data \cite{info:doi/10.2196/55779}. It enables semantic modeling of clinical concepts through archetypes and templates, ensuring both syntactic and semantic consistency of health data. However, integration of openEHR-based systems into analytical environments such as OMOP CDM still requires dedicated transformation logic to map archetype content to OMOP tables and terminologies.

To address this challenge, we previously introduced the Eos engine and the OMOP Conversion Language (OMOCL), an open-source transformation framework and associated mapping language that enable the conversion of structured clinical data from openEHR into the OMOP CDM using archetype-driven mappings \cite{kohler_eos_2023}. Eos and OMOCL demonstrated the feasibility of an automated, standards-based ETL process, leveraging openEHR archetype semantics for reproducible data transformation.

While these initial results confirmed the technical viability of the approach, several practical limitations became apparent. In the previous implementation, internal value sets within archetypes were only partially supported. The visit-generation logic could not be easily adapted to local data structures. In addition, the international mapping library covered only a limited subset of the archetypes available in the openEHR Clinical Knowledge Manager (CKM). Moreover, our earlier work revealed that several limitations stemmed not from the transformation framework, but from structural constraints within the OMOP CDM. Structural restrictions and the lack of expressive mechanisms for representing contextual clinical information led to unavoidable semantic loss during transformation. These findings necessitated a further investigation into the alignment between openEHR archetypes and OMOP's representational capabilities. These constraints limited both the completeness and reusability of the generated mappings, reducing the framework's potential for broader adoption.

The present work aims to implement a new generation of the Eos and OMOCL framework to address these critical limitations and further advance openEHR-to-OMOP interoperability. Concretely, it contributes a mechanism for mapping internal openEHR value sets through \textit{conceptMaps}, an Archetype Query Language (AQL)-driven approach for flexible visit occurrence generation, an expanded, systematically curated library of archetype mappings, and a systematic analysis of the semantic and structural limitations of the OMOP CDM that these mappings reveal. As part of this we assessed OMOP CDM's capability to represent data from these 198 archetypes, evaluating mapping coverage, terminology completeness, domain distribution, and identifying representational gaps.

By improving the amount and quality of archetype-to-OMOP mappings, these new capabilities enable more comprehensive and semantically accurate transformation of openEHR data for secondary use. This in turn facilitates the reproducible integration of clinical and research datasets, promoting standards-based interoperability and supporting evidence-based decision-making across healthcare and research domains.

In our previous work, we had already identified several structural and semantic limitations within the OMOP CDM that affected the completeness of openEHR-to-OMOP transformations. Building on these initial observations, the expanded mapping effort in this study provides an opportunity to systematically analyze these limitations across a substantially larger set of archetypes mapped to OMOP. This detailed evaluation aims to reveal where OMOP currently lacks representational depth for research use, particularly in areas where granular clinical information cannot be mapped without 
semantic loss that might impact clinical studies.

\section{Material and methods}

\subsection{OMOP Conversion Language (OMOCL)}
OMOCL is a domain-specific language (DSL) designed to provide a human-readable and machine-executable format for defining mappings from openEHR to the OMOP Common Data Model (CDM). It enables the transformation of openEHR records into the format required by OMOP.

OMOCL mappings are archetype-driven: each openEHR archetype is mapped to one or more OMOP CDM tables. Within these mappings, specific data elements from the archetype, identified by their unique archetype path, are aligned with corresponding fields in the OMOP CDM table. These path-to-field mappings capture how the individual data elements of clinical observations or measurements are transformed. Being based on archetypes allows for a high reusability of mappings. OMOCL is an open specification and also comes with an international library of open-source mappings from openEHR to OMOP, accessible on GitHub \cite{OMOCL2025}.

Although the initial version of OMOCL demonstrated the feasibility of archetype-based transformation, it lacked the functionality to map internal openEHR value sets. Additionally, the international mapping library remains incomplete, limiting reuse across clinical domains. These gaps present challenges in scaling transformation pipelines across heterogeneous EHR environments. In this work, we introduce enhancements to OMOCL that address its limitations and significantly extend the mapping library \cite{kohler_eos_2023}.

\subsection{Eos}
Eos is an ETL engine designed to automate the transformation of clinical data from openEHR into the OMOP CDM by executing mappings defined in OMOCL. Implemented as an application server with a structured API, Eos loads openEHR compositions, applies OMOCL-defined transformations, and populates the corresponding OMOP CDM tables. This automated process reduces the need for manual intervention, minimizes transformation errors, and supports reproducible research by enabling consistent and standardized data harmonization across heterogeneous clinical data sources \cite{kohler_eos_2023}. Eos is open source and accessible on GitHub \cite{Eos2025}.

The Eos engine provides the first successful open-source implementation capable of processing OMOCL for transforming openEHR records into OMOP. A functional gap remains in the generation of visit\_occurrence records. In OMOP, this table represents a healthcare encounter or stay. Eos currently generates visit\_occurrences on a per-composition basis and does not yet support visits transformed from openEHR templates, which limits meaningful encounter groupings.

To address these limitations, we developed the engine to enable visit\_occurrence generation through AQL queries. We also enhanced Eos to support the extended functionalities introduced in OMOCL.

\subsection{Method for Analyzing OMOP Limitations}

To assess structural and semantic limitations of the OMOP CDM, we analyzed the mappings of 198 openEHR archetypes to OMOP tables. The unit of analysis was the archetype-specific OMOCL mapping definition, with particular focus on primary OMOP concept identifier fields (e.g., condition\_concept\_id, measurement\_concept\_id). A primary concept\_id is the field that identifies the record itself, which the CDM recommends for primary use in analyses \cite{CDM54}. We conducted this as a directed content analysis. From an initial pass over the mapping library we derived a codebook of recurring limitations. Three of these could be measured directly across all mappings and form the quantitative core of the analysis, vocabulary gaps, structural fragmentation, and semantic field overloading, as summarised in Table \ref{codebook}.

\begin{table}[h]
    \centering
    \caption{The three measured OMOP limitation categories}
    \label{codebook}
    \begin{tabular}{|p{0.4cm}|p{3.0cm}|p{8.0cm}|}
    \hline
    \textbf{\#} & \textbf{Category} & \textbf{What it is} \\
    \hline
    1a & Vocabulary gap & Mapped field has no standard OMOP concept \\
    \hline
    1b & Vocabulary gap & Internal openEHR value code has no standard OMOP concept \\
    \hline
    2 & Fragmentation & One archetype dispersed across many OMOP table instances \\
    \hline
    3 & Field overloading & One OMOP column fed by several distinct source paths, a lossy collapse of several sources into one column \\
    \hline
    \end{tabular}
    {\small The three categories were defined in advance and quantified across all 198 mappings. The measurement rules and} further qualitative limitations are described in the text.
\end{table}

The three categories were quantified across all 198 archetypes using deterministic rules that operate on the mappings and the underlying archetypes. Vocabulary gaps were measured at two levels. At the concept level, a gap was recorded whenever the primary concept identifier field of a table could not be resolved to a standard OMOP concept and therefore defaulted to concept 0. At the value level, each conceptMap entry that could not be mapped to an OMOP concept was recorded. Both levels are properties of the mappings and not of a particular dataset. They rest on different denominators and are therefore reported separately and not pooled.

Structural fragmentation was measured directly from the mappings. For each archetype we compared its nodes against the OMOP table instance they are mapped to. A node was counted as fragmented when the mapping did not write it into the primary instance, because it populated another table instance. If mapped to another table it requires an auxiliary record linked back through a fact\_relationship. A cluster slot was counted as one such link, and the cluster itself was then compared against its own table in the same way. The measure counts nodes with no field in the target table. It does not distinguish a node that could be carried by an auxiliary record from one OMOP cannot represent at all. Separating them would require a concept by concept assessment of 1911 nodes against the full OMOP vocabulary, which is a separate study. A node OMOP can represent still requires an auxiliary record, so the count does not depend on which case applies. It also captures any field a mapping author may have missed. The mappings undergo community peer review, so we assume such omissions to be rare. To compare archetypes and tables we counted semantically identical OMOP fields as one. Such fields are different representations of a single item, for example a concept and its source concept, or the string, numeric and concept variants that carry the value. A single openEHR node can likewise cover several OMOP fields. A DV\_QUANTITY, for example, carries both the value and its unit, which OMOP holds in two separate fields. We merged such fields only where the OMOCL mapping read them from the same archetype path, since that is what shows a single node feeds both. We counted only the nodes an archetype itself models, so the further context fields carried by the reference model were left out. Quantifying that context was beyond the scope of this work. Their omission makes the measure conservative, since counting them would raise the fragmentation and not lower it.

Semantic field overloading was measured by counting the OMOP columns into which the mapping converged several archetype paths. A column was counted as overloaded when at least one mapping fed it from two or more distinct source paths. The denominator is the 87 columns the library writes into, out of the 175 columns available in the ten tables it targets. Columns that no mapping writes were excluded. Fields filled from a constant value and conceptMap entries were excluded as well. Paths carried by the reference model were retained, since an event time and a context time are distinct timestamps whose convergence on one column is the effect under measurement. Where OMOP provides only one value, a single path was mapped and the remaining paths were left unmapped. These cases were not counted as overloading.

The openEHR Problem/Diagnosis archetype was analyzed in detail as a worked example, selected due to its central role in clinical documentation and research. It illustrates the structural categories in practice, in particular the fragmentation of a single clinical statement into auxiliary Observation and fact\_relationship records.

We carried out the implementation of the analytics of the mappings with the assistance of coding agents (Claude Opus 4.8, Anthropic) to speed up the process. The authors checked and verified all results and take full responsibility for the analysis. The full code is openly available in a public repository for transparency \cite{OMOCLanalytics2025}.


\section{Results}

\subsection{ConceptMaps}
\begin{figure}[h]
    \centering
     \caption{GOSE archetype category element}
    \label{Glasgow}
     \includegraphics[width=1\linewidth]{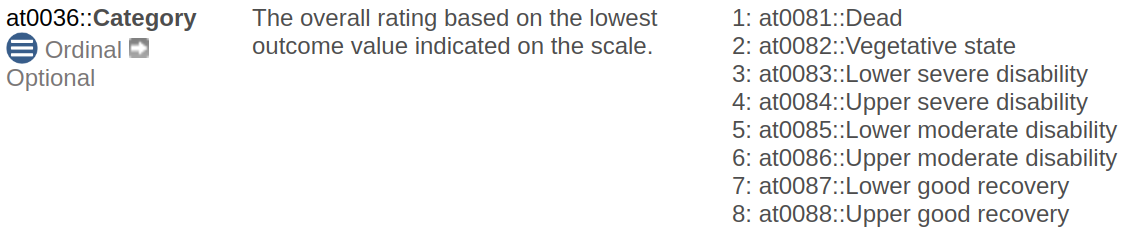}
    {\small
    Glasgow Outcome Scale--Extended (GOSE) archetype: category element illustrating internal value set coding.
    }
\end{figure}

OpenEHR archetypes usually contain rich internal terminology which can define state-of-the-art value sets (scales, grades, etc.). These are internally coded with at-codes. As an example, Figure \ref{Glasgow} shows the Glasgow Outcome Scale category Element. For the transformation into OMOP, these codes need to be mapped to a terminology that is integrated into the OMOP vocabulary.

\begin{figure}[h]
    \centering
    \caption{Example of conceptMap mapping}
    \label{conceptMap}
    \includegraphics[width=\linewidth]{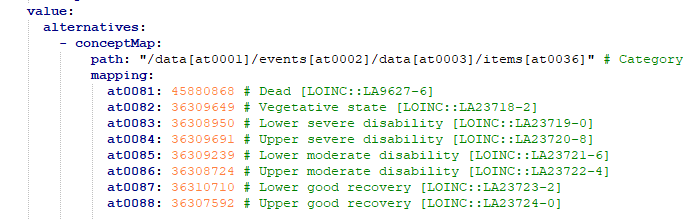}
    {\small Snippet of an OMOCL mapping showing how internal openEHR codes are linked to OMOP standard concept identifiers using the conceptMap mechanism.}
\end{figure}

Making use of these internal value sets provides invaluable knowledge to the target OMOP data, specifically for Measurements and Observations. We added support for mapping internal archetype codes via a new logic called conceptMaps. Hereby, the internal at-codes of an archetype are mapped to concept ids in OMOP. For each of the at-codes, a corresponding concept id in OMOP is mapped. An example for a conceptMap mapping of the internal coding of the GOSE archetype is illustrated in Figure \ref{conceptMap}.

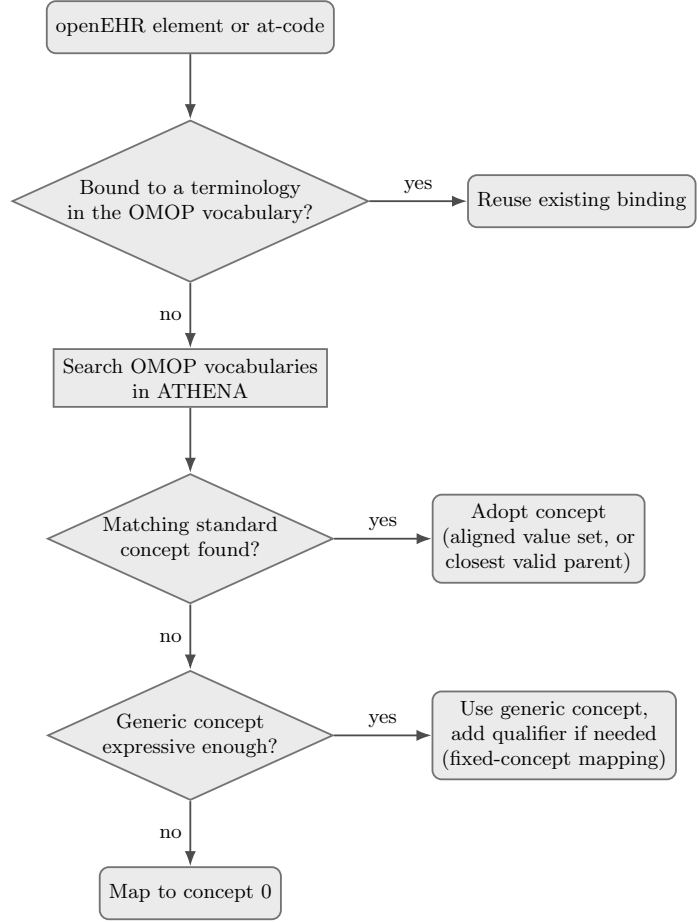
\begin{figure}[h]
    \centering 
    \caption{Concept identification procedure}
   \resizebox{0.7\linewidth}{!}{ \begin{tikzpicture}[
        node distance=1.0cm and 1.5cm,
        every node/.style={font=\small},
        startstop/.style={rectangle, rounded corners, draw=black!55, fill=black!8, thick, align=center, minimum height=0.8cm, inner sep=4pt},
        process/.style={rectangle, draw=black!55, fill=black!8, thick, align=center, minimum height=0.8cm, inner sep=4pt},
        decision/.style={diamond, aspect=2.2, draw=black!55, fill=black!8, thick, align=center, inner sep=1pt},
        result/.style={rectangle, rounded corners, draw=black!55, fill=black!8, thick, align=center, minimum height=0.8cm, inner sep=4pt},
        arrow/.style={-{Latex}, thick, draw=black!70}
    ]
    \node[startstop] (start) {openEHR element or at-code};
    \node[decision, below=of start] (d1) {Bound to a terminology\\in the OMOP vocabulary?};
    \node[result, right=of d1] (reuse) {Reuse existing binding};
    \node[process, below=of d1] (search) {Search OMOP vocabularies\\in ATHENA};
    \node[decision, below=of search] (d2) {Matching standard\\concept found?};
    \node[result, right=of d2] (adopt) {Adopt concept\\(aligned value set, or\\closest valid parent)};
    \node[decision, below=of d2] (d3) {Generic concept\\expressive enough?};
    \node[result, right=of d3] (generic) {Use generic concept,\\add qualifier if needed\\(fixed-concept mapping)};
    \node[result, below=of d3] (zero) {Map to concept 0};

    \draw[arrow] (start) -- (d1);
    \draw[arrow] (d1) -- node[above]{yes} (reuse);
    \draw[arrow] (d1) -- node[left]{no} (search);
    \draw[arrow] (search) -- (d2);
    \draw[arrow] (d2) -- node[above]{yes} (adopt);
    \draw[arrow] (d2) -- node[left]{no} (d3);
    \draw[arrow] (d3) -- node[above]{yes} (generic);
    \draw[arrow] (d3) -- node[left]{no} (zero);
    \end{tikzpicture}}
     {\small
    
    Procedure for identifying the OMOP concept of a mapping. conceptMaps translate internal at-codes to OMOP concepts as a one-to-one mapping. When no dedicated concept exists, a more general concept is used if it is expressive enough, otherwise the value is mapped to concept 0.
    }
    \label{conceptflow}
\end{figure}

Domain experts identified the OMOP concepts for the mappings manually, following the procedure summarized in Figure \ref{conceptflow}. For conceptMaps, they applied it to each internal at-code. When an at-code was already bound to an external terminology that is part of the OMOP vocabulary, that binding was reused. When no binding existed, the OMOP vocabularies were searched manually with the OHDSI ATHENA browser for a standard concept that matched the at-code and its textual description. If the terminology defined a value set of answers that aligned with the openEHR codes, that value set was adopted. When no suitable concept was found, the at-code was mapped to concept 0.

Each at-code was mapped to a single OMOP concept, so the conceptMaps are one-to-one. openEHR value sets are sometimes more granular than the available OMOP concepts. In these cases we mapped to the closest parent concept that remained semantically valid. For example, in the New York Heart Association heart failure scale the openEHR subclasses Class IIIa and Class IIIb were both mapped to concept 36307909 for Class III, because the terminology has no code for these subtypes.

The same concept identification applied to the mappings in general. When no dedicated OMOP concept existed for an element, we used a more general concept. Where needed, we added a qualifier concept in the same record to keep the clinical meaning. For example, the MSFC score has no dedicated OMOP concept. It is represented through a fixed-concept mapping to the generic concept 4175713 for a severity score, qualified by concept 374919 for multiple sclerosis. This introduced some semantic loss and was applied only when the general concept still captured the clinical meaning.

As a result, conceptMaps improve the semantics being transformed into OMOP from openEHR and lower the amount of data lost when integrating, as these at-codes are typically not coded in the international Clinical Knowledge Manager (CKM) \cite{CKM2025} and would therefore be assigned the unknown concept code (0).

In the new OMOCL generation, we identified and added such conceptMaps to the library. It contains 55 conceptMaps across 37 archetypes, holding 283 entries in total. Of these, 35 translate a value, 11 a qualifier, and 9 concept\_ids.

The conceptMaps and the mappings in general are part of the open OMOCL library and undergo community peer review. We are working to broaden this review by involving more reviewers and to establish governance for the mapping library.

\subsection{Visit occurrence}
One of the key elements in OMOP are the visits, which are related to every table that contains clinical information (Condition\_occurrence, Procedure\_occurrence, Drug\_exposure, Measurement, Observation, etc.). The initial approach of Eos for the visit\_occurrence table generation was to generate a visit for each composition in openEHR. This is usually correct as each composition contains the context to provide start and end dates of a given clinical act.

While this contains the start and end date of a clinical act, it covers only a subset of the overall hospital encounter, from admission to discharge. To cover this encounter, some openEHR systems implement this at the template level. Therefore, there is a need to generate visits directly from the composition or context content. For this use case, we improved Eos to be able to query the openEHR system with an AQL and populate the visit\_occurrence from that.

AQL is executed over the EHR repository, retrieving four different values from the query.
\begin{itemize}
    \item \textbf{EHR ID:} The ehr\_id from the patient.
    \item \textbf{Source visit ID:} The identifier for a given visit, regardless of source data type.
    \item \textbf{Start date:} Start date of a given source visit.
    \item \textbf{End date:} End date of a given source visit.
\end{itemize}

This process groups all visits by EHR\_ID and source visit id, calculating the minimum start date and maximum end date for all the obtained values. This ensures that visits will span the full duration of the source visit, even if different times exist in the EHR. Leveraging AQL as the foundation for visit\_occurrence generation offers fine-grained control over the construction of visits. This is enabled by AQL's ability to query not only composition data and contextual attributes, but also hierarchical structures such as Folders and administrative metadata including the EHR\_STATUS object. As a result, this improves the adaptability of Eos to openEHR platforms.

\subsection{Extending and evaluating the international mapping library}
As part of a community-driven effort, we extended the OMOCL library to cover nearly all published archetypes, resulting in 198 archetype mappings (including two deprecated and one under review status). The deprecated and under-review archetypes were included because they were present in the initial test datasets and required transformation support. This expansion significantly enhances the library's coverage and usability for standardizing clinical data into the OMOP Common Data Model. It facilitates interoperability across health information systems, reduces duplication of mapping efforts, and accelerates research by enabling seamless integration of openEHR-based data into the OMOP ecosystem.

\begin{figure}
    \centering
     \caption{Distribution of mapped archetypes across OMOP CDM domains}
    \label{pieChart}
    \includegraphics[width=1\linewidth]{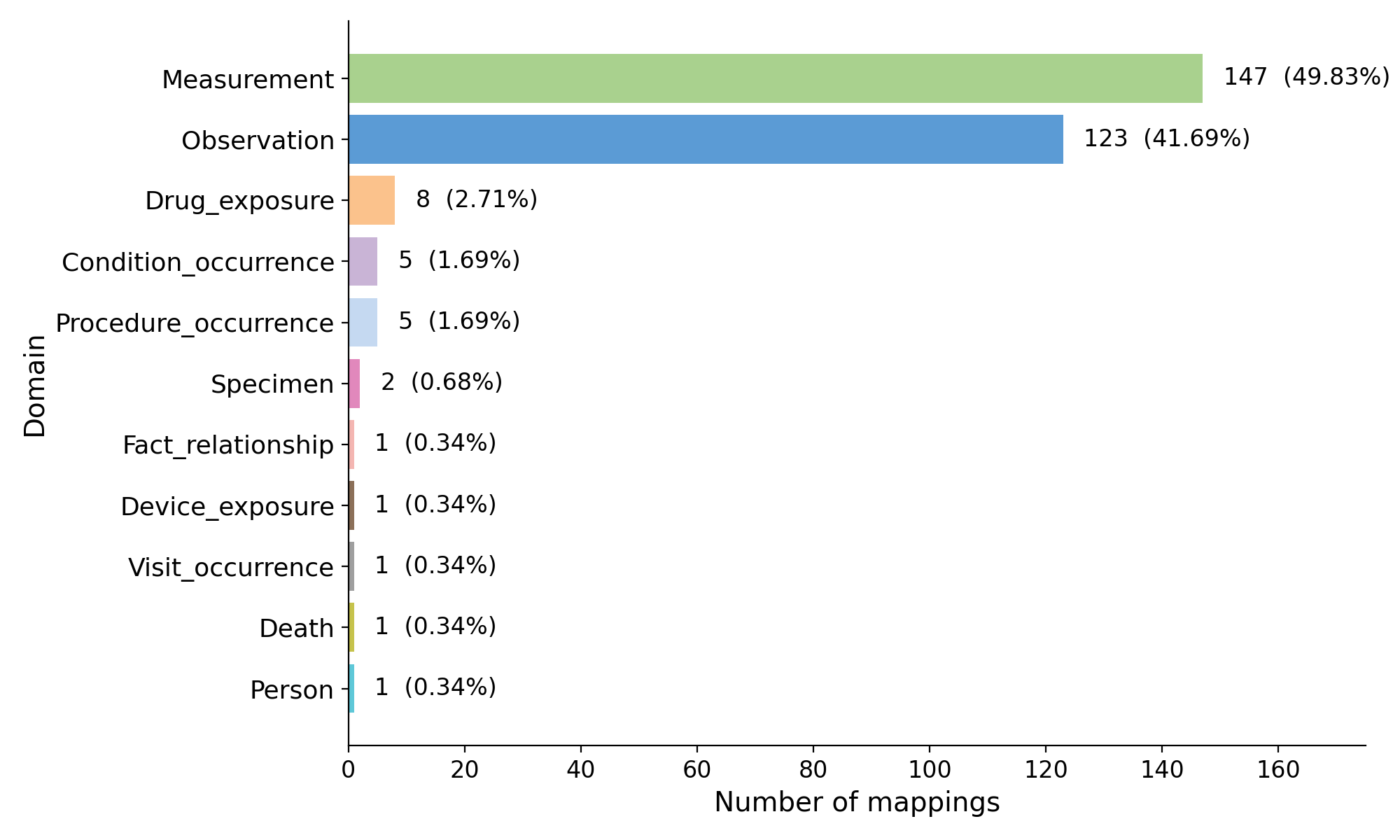}
      {\small A total of 295 entries were mapped to standard OMOP CDM tables. The largest proportions were found in Measurement (147 mappings, 49.83\%) and Observation (123 mappings, 41.69\%). Smaller contributions came from Drug\_exposure (8 mappings, 2.71\%), Condition\_occurrence (5 mappings, 1.69\%), Procedure\_occurrence (5 mappings, 1.69\%), Fact\_relationship (1 mapping, 0.34\%), Device\_exposure (1 mapping, 0.34\%), Specimen (2 mappings, 0.68\%), Visit\_occurrence (1 mapping, 0.34\%), Death (1 mapping, 0.34\%), and Person (1 mapping, 0.34\%).}
\end{figure}

Around 91\% of the mapped entries target the Observation (41.7\%) and Measurement (49.8\%) tables, illustrated in Figure \ref{pieChart}. Meanwhile, the other domains do not cover as many archetypes. This shows that openEHR modeling focuses substantially on what OMOP identifies as Observations and Measurement. It may indicate either that many archetypes describe clinical findings and quantitative data, or that these OMOP domains function as structurally broad target tables for heterogeneous clinical concepts. For the other domains, there seem to be a few established archetypes that cover most use cases.

This distribution follows from how the two models are structured. In OMOP the target domain is not chosen by the mapper but follows from the standard concept, whose domain\_id assigns each element to a table. A diagnosis does not require a new archetype for every condition, the single Problem/Diagnosis archetype represents them all and the specific condition is supplied by the bound terminology in the data and the usage of clusters. The same holds for procedures. Measurements and observations are more diverse. Each new type of finding or measurement introduces different elements and different semantics that existing archetypes do not capture, so their number keeps growing as clinical research adds new instruments and scores. This is why one archetype maps to Condition\_occurrence or Procedure\_occurrence while many distinct archetypes resolve to Measurement and Observation. The same pattern appears in FHIR, where a single Condition resource stands against a large and continually extended family of Observation profiles and extensions \cite{kramer_fhir_2022}.

Not all published archetypes were mapped. Composition archetypes were not mapped since these merely act as containers. Ten other published archetypes could not be mapped to specific OMOP concepts or tables for various reasons, for a detailed description see Appendix \ref{tab:archetypes_no_mapping}.

The systematic gap analysis described in the methodology was applied to all 198 archetype mappings. The results are illustrated in Table \ref{fragmentation}. Vocabulary gaps left 8.59\% of the primary concept identifiers and 7.1\% of the conceptMap entries without a standard OMOP concept. Structural fragmentation affected 87.2\% of the openEHR nodes, which have no direct field in their OMOP table instance. Semantic field overloading was recorded for 35 of the 87 OMOP columns the library writes into (40.2\%). We describe each category in turn.

\begin{table}[h]
    \centering
    \caption{Summary of the three measured OMOP limitation categories}
    \label{fragmentation}
    \renewcommand{\arraystretch}{1.6}
    \setlength{\tabcolsep}{8pt}
    \begin{tabular}{|c|>{\raggedright\arraybackslash}p{2.3cm}|>{\raggedright\arraybackslash}p{5.9cm}|c|}
    \hline
    \textbf{\#} & \textbf{Category} & \textbf{Measure} & \textbf{Result} \\
    \hline
    1a & Vocabulary gap & Primary concept\_id fields resolving to concept 0 & 25 of 291 (8.59\%) \\
    \hline
    1b & Vocabulary gap & conceptMap entries resolving to concept 0 & 20 of 283 (7.1\%) \\
    \hline
    2 & Fragmentation & openEHR nodes without a direct field in the target OMOP table instance & 1996 of 2289 (87.2\%) \\
    \hline
    3 & Field overloading & CDM columns fed by more than one distinct source path & 35 of 87 (40.2\%) \\
    \hline
    \end{tabular}
    {\small Categories and their numbering follow Table \ref{codebook}. Category 1a is measured over the primary concept\_id mapping blocks of Table \ref{mappingproperties}, category 1b over the individual conceptMap entries. Category 2 is based on 2289 nodes across 198 archetype mappings, category 3 on the 87 CDM columns the library writes into. }
\end{table}

The frequently encountered vocabulary gaps in OMOP hindered the transformation and limited concept mapping for clinical data elements. Table \ref{mappingproperties} shows in detail how the primary OMOP concept\_id was determined for each mapping. The analysis covered only the primary OMOP concept\_id fields (e.g., measurement\_concept\_id), as these are recommended for primary use in analyses. Four categories of mappings were identified. A fixed concept mapping assigns a constant OMOP concept ID. For example the height archetype always maps to concept 3036277 for body height. A mapping with paths reads the concept\_id from the data contained in the archetype node, such as the diagnosis name containing the terminology code. A mapping with a conceptMap translates internal openEHR at-codes to OMOP concepts, such as the Glasgow Outcome Scale categories. The fourth category, zero, applies when no corresponding OMOP concept exists, and therefore no mapping can be performed. This revealed that 8.59\% of the fields mapped from openEHR records into the OMOP CDM could not be mapped to a terminology code in OMOP and were therefore assigned to this "No matching concept", concept\_id 0. 

\begin{table}[h]
    \centering
    \caption{How the primary OMOP concept\_id is determined}
    \label{mappingproperties}
    \begin{tabular}{|l|r|r|}
    \hline
    \textbf{Property / Concept ID Type} & \textbf{Count} & \textbf{Percentage} \\
    \hline
    Mappings to zero (no applicable concept) & 25 & 8.59\% \\
    Mappings using a fixed concept & 234 & 80.41\% \\
    Mappings with paths & 24 & 8.25\% \\
    Mappings with conceptMap & 9 & 3.09\% \\
    \hline
    \end{tabular}
    {\small
    Counts are based on unique primary OMOP concept\_id mapping blocks, with each block counted once.
    Properties are not mutually exclusive, a single mapping block may exhibit multiple properties.
    Percentages refer to the proportion of mapping blocks showing the respective property. }
\end{table}

The conceptMaps were measured separately. Of the 283 conceptMap entries in the library, 20 (7.1\%) translate an internal at-code to concept 0, leaving the openEHR value without an OMOP counterpart. This proportion is a property of the mappings themselves. How strongly a given dataset is affected additionally depends on how often the individual at-codes occur in it. Neither vocabulary gap measurement accounts for cases where an expressive enough generic term was available and used instead of a more specific one.

Several archetype elements required auxiliary Observation or Fact Relationship records due to structural mismatches between openEHR and OMOP. For example, the body site of a diagnosis has no dedicated field in the condition\_occurrence table of OMOP. To represent it, it must be stored as an Observation that is linked back to the condition. This linkage is done in OMOP through a fact\_relationship entry. Such fragmentation is present across the library. Only 12.8\% of the openEHR nodes have a direct field in their OMOP table instance. The remaining 87.2\% have none. These nodes need an auxiliary record of this kind, unless OMOP cannot represent them at all. Fragmentation is spread across the library and not confined to a few large archetypes. In 187 of the 198 archetypes (94.4\%), at least one node falls outside the target table. Archetype nodes counted against the content fields of their target table set a floor of 60.4\%, well below the measured fragmentation. The floor assumes that any node could occupy any field, which OMOP's typed columns do not allow, so it can only understate the fragmentation. It shows that the greater part of the fragmentation follows from the size difference between the models alone and cannot be attributed to our mapping decisions. We implemented the use of the fact\_relationship table as part of a programmed CustomMapping process \cite{kohler_eos_2023}. While a significantly higher number of fact\_relationship mappings would be expected, their implementation was limited due to insufficient support in existing OMOP tooling. It is important to note, however, that this analysis is based solely on the currently published archetypes, additional archetypes in development or used locally may further impact domain distribution in the future.

The library writes into 87 CDM columns, and 35 of them (40.2\%) are overloaded in at least one mapping. Twenty one of these (60\%) are dates or datetimes, where one column receives several clinically distinct timestamps, since each OMOP record exposes a single timestamp per role. The remaining fourteen (40\%) are more severe and affect value, qualifier and concept identifier columns, where the column carries a different meaning from one record to the next. This count is a lower bound, because only columns that a mapping feeds from several sources are counted. A column overloaded by design but mapped from the single source considered correct appears clean, for example condition\_status\_concept\_id, which carries several distinct diagnostic qualifiers.


\section{Detailed Analysis of OMOP Limitations}

\subsection{Fact relationships}
When mapping openEHR archetypes to the OMOP CDM, many data elements cannot be stored in the target table itself. These elements must instead be represented in separate OMOP tables like the Observation table and then linked back to the primary record using the fact\_relationship table. While event fields from the measurement and observation tables could be used for this relationship, the relation semantics between the fields are not explicitly stated and therefore cannot be used in federated research. The relationship\_concept\_id field specifies how two records are related and is therefore critical for preserving the original clinical meaning.

This mechanism allows OMOP tables with a limited number of fields to be extended indirectly. However, it introduces important limitations. Federated analytics workflows must be explicitly designed to traverse fact\_relationship links in order to reconstruct a complete clinical entity. Furthermore, correct interpretation depends on the availability of appropriate relationship\_concept\_id values. If such relationship concepts are missing or overly generic, the semantics of the link must be inferred from the linked record itself rather than from the relationship, increasing query complexity and reducing interoperability.

\begin{table}[h!]
    \centering
    \caption{Mapping openEHR Problem/Diagnosis to OMOP, limitations and combined Observation \& Fact Relationship requirements}

    \label{diagnosismapping}
    \begin{tabular}{|l|p{4cm}|p{4cm}|}
        \hline
        \textbf{openEHR field} &
        \textbf{OMOP target} &
        \textbf{Limitation} \\
        \hline

        Problem / Diagnosis name & condition\_occurrence.
        condition\_concept\_id &
        -  \\
        \hline

        Variant &
        observation. value\_as\_concept\_id &
        Requires observation and fact\_relationship link. \\
        \hline

        Clinical description &
        observation. value\_as\_string &
        Requires observation and fact\_relationship link. \\
        \hline

        Body site &
        observation. value\_as\_concept\_id &
        Requires observation and fact\_relationship link. Could be stored as part of the condition code. \\
        \hline

        Structured body site &
        observation. value\_as\_concept\_id & Cluster used here usually in openEHR is openEHR anatomical\_location cluster. Which results in multiple observation and fact\_relationship links.  \\
        \hline

        Cause / Aetiology &
        observation. value\_as\_concept\_id &
        Requires observation and fact\_relationship link. \\
        \hline

        Date/time of onset & condition\_occurrence.
        condition\_start\_date &
        -  \\
        \hline

        Clinically recognised time &  - &
        Diagnostic timing lost, no dedicated field.  \\
        \hline

        Severity &
        observation. value\_as\_concept\_id &
        Requires observation and fact\_relationship link. \\
        \hline

        Specific details &
        observation. value\_as\_string &
        Cluster used here. Which results
        in multiple
        fact relationship links. \\
        \hline

        Course description &
        observation. value\_as\_string &
        Requires observation and fact\_relationship link.  \\
        \hline

        Impact &
        observation. value\_as\_concept\_id &
        Requires observation and fact\_relationship link. \\
        \hline

        Date/time of resolution & condition\_occurrence.
        condition\_end\_date &
        -  \\
        \hline

        Status &
        condition\_occurrence.  condition\_status\_concept\_id &
        - \\
        \hline

        Diagnostic certainty &
        observation. value\_as\_concept\_id &
        Requires observation and fact\_relationship link. \\
        \hline

        Comment &
        note. text &
        note and fact\_relationship link required.  \\
        \hline
    \end{tabular}
\end{table}

The fragmentation measure shows that these structural limitation patterns hold across the whole mapping library. To illustrate their practical impact, we analysed the diagnosis archetype in full detail, as diagnosis represents a fundamental clinical concept that illustrates the structural challenges. As shown in Table \ref{diagnosismapping} only 4 of 16 nodes could be directly mapped to the condition\_occurrence table. The clinically recognized time could not be directly represented. OMOP provides no dedicated datetime field in the condition\_occurrence table for this attribute. Modeling it as an observation would require a standard concept representing "clinically recognized time," which is currently not available in the OMOP vocabularies. The other 11 each require a record outside condition\_occurrence. Ten become observations and the comment becomes a note. All 11 are linked back to the condition through a fact\_relationship. Two of the ten observations are CLUSTER fields, each counted here as a single anchor observation. If these clusters are expanded, the number of records and links increases significantly. This breakdown was derived by hand and matches the automated fragmentation measure, which counts 4 directly mapped nodes and 12 auxiliary nodes for this archetype. That measure counts every element the mapping does not place anywhere as fragmentation. The clinically recognized time is such an element, because no OMOP concept is available for it. The count of 12 is therefore correct but not precise. It does not distinguish an element that requires an auxiliary record from one that cannot be represented at all.

Only one of the required fact\_relationship entries, the aetiology link, has a matching standard relationship\_concept\_id in OMOP. The others have none. For example, the body site of a diagnosis must be stored as a separate Observation linked back to the condition, but OMOP provides no standard relationship concept for such a link. By convention, \_concept\_id fields must reference standard concepts \cite{CDM54}. If this convention is disregarded, there is a non-standard, OMOP-generated relationship concept   "Finding site of", but it semantically describes relationships between vocabulary concepts and not between data records and therefore does not properly fit.  In these cases a zero relationship\_concept\_id would need to be used. The semantic meaning of each link would therefore depend entirely on the concept identifiers contained in the linked Observation record rather than on the relationship code itself. This complicates the use of these relationships in federated analytics because queries must not only navigate the links but must also infer the nature of the relationship from the target record.

Introducing specific relationship\_concept\_id values for all possible clinical relations is not feasible. The number of required relationships would depend on the granularity of the modeled qualifiers and would expand into thousands of potential combinations. For comparison, OMOP already contains approximately 744 relationship concepts used in the social context domain, such as family relationships (e.g. Mother) \cite{Athena}. Extending this approach to all clinical qualifiers would require an entire vocabulary of clinical relationship concepts. The alternative is to fall back to a zero relationship\_concept\_id, which provides very limited semantic value. Any user who needs to identify specific types of clinical relationships would need to resolve both the relationship code and the concept identifier of the linked Observation, adding an additional abstraction layer to each query.

Archetypes allow specifying more detailed data structures for more detailed information needs, in the form of CLUSTERS. When these structures are used for detailed data needs, the number of fact\_relationship links increases further. For example, the cluster problem\_qualifier, which records context specific or time specific diagnostic qualifiers, contains multiple individual data elements. A mapping table can be found under Appendix table \ref{tab:problem_qualifier_factrel}. Each element requires its own fact\_relationship link entry. As a result, the complexity and number of linked OMOP tables scale, and the representation of a single diagnosis becomes a large constellation of loosely connected records rather than a cohesive clinical entity.

\subsection{Terminology-Driven (TDCI) vs. Model-Driven Concept Identification (MDCI)}
All but one of the fact\_relationship links reference observation records. This shows that OMOP relies heavily on observation to store additional clinical information. This causes semantically distinct elements to be aggregated into a single table type, which substantially complicates data access. For a diagnosis archetype, this already produces 10 observation entries (Table \ref{diagnosismapping}). When combining both the diagnosis and problem\_qualifier archetypes, this number rises to 23, ten observations from the diagnosis archetype plus the thirteen observation rows of the problem\_qualifier cluster (Appendix Table \ref{tab:problem_qualifier_factrel}), not including the two additional cluster slots that would further increase the count. All of these are linked via fact\_relationships. Apart from that also 41.7\% of all current mappings from openEHR archetypes already result in observations.

The OMOP CDM is based, similar to FHIR, on  the Terminology-Driven Concept Identification (TDCI) approach. This means that the clinical concept is defined by the terminology code. The model is generic, and codes provide the meaning, users must locate the appropriate concept from terminologies. In Table \ref{tab:Measurement} this is illustrated. The Measurement table itself is generic for all measurements, and semantics are added by the terminology identifying this concept. If users want to access this information they have to query the correct code.

The TDCI approach offers significant flexibility by relying on terminology codes to assign meaning to generic model structures, allowing new clinical concepts to be represented without modifying the underlying data model. This makes it well suited for data exchange scenarios, where a shared terminology codings provides semantic alignment, and for analytical use cases such as OMOP, where a flat denormalized structure enables efficient large-scale cohort queries across sites.

\begin{table}[h]
    \caption{Example of TDCI using the OMOP  table Measurement depicting a systolic blood pressure}
    \centering
    \begin{tabular}{|c|c|c|}
        \hline
        \textbf{field name} & value & description \\
        \hline
        concept id & 271649006 (SNOMED code) & identifies what clinical concept is represented  \\
        value\_as\_number & 180  & contains the value connected to this concept\\
        unit\_concept\_id & mmHg & unit connected to this concept \\
        \hline
    \end{tabular}
    \label{tab:Measurement}
\end{table}

openEHR uses a Model-Driven Concept Identification (MDCI) approach, meaning that the clinical concept is defined and identified by the model itself. An archetype represents an explicit clinical concept whose meaning is specified within the model, and each data element inside the archetype has its semantics defined as part of that structure. This is illustrated in Table \ref{tab:bparche}. If users want to access this information they have to query the model instead of a code. Additionally terminologies in openEHR can be annotated to the fields, and should be, but the interoperability of the standard does not fundamentally require those. This distinction refers to how concepts are identified in the model, not to the use of terminology codes as data values (e.g., an ICD code for a diagnosis), which is required in both approaches. The usage of terminology for this is necessary in both TDCI and MDCI.

\begin{table}
    \caption{Example for MDCI from openEHR using the blood pressure archetype}
    \centering
    \begin{tabular}{|c|c|p{5cm}|}
        \hline
        \textbf{field name} & value & description \\
        \hline
        systolic blood pressure & 180 mmHg & element for systolic blood pressure, of type quantity having a magnitude and unit. \\
        \hline
    \end{tabular}
    \label{tab:bparche}
\end{table}

The approaches are not mutually exclusive, standards employ terminology codes and structured fields to varying degrees. The distinction lies in what predominantly carries clinical meaning, in OMOP and FHIR, generic structures are given meaning primarily through terminology codes, whereas in openEHR, clinical meaning is predominantly encoded in the model itself, with terminology codes serving as an additional semantic layer.

The TDCI approach has limitations. The clinical concept is identified through a terminology code, therefore, all users must select the same code from the same terminology in order to represent that concept consistently across sites. This requires strong governance to ensure uniform coding practices. In addition, as more clinical concepts are represented across multiple OMOP tables, the effort required to identify the correct codes and assemble the relevant data increases substantially, making data access and interpretation progressively more complex.

This presupposes that an appropriate code exists in the reference terminology. When no suitable standard code is available, implementers must rely on local or custom codes, which causes variation in how the same concept is represented. This limitation appears in our mapping results, where 8.59 percent of the primary concept\_id fields in the main OMOP tables had no matching OMOP standard concept and were assigned "No matching concept", concept\_id 0. At the value level, 7.1 percent of the conceptMap entries had no OMOP counterpart either. Mapping additional contextual or metadata fields by using tables such as observation or fact\_relationship would likely increase the proportion of unmapped values, because each of these elements requires its own terminology binding. As the number of such elements grows, vocabulary gaps become more visible and semantic completeness and interoperability decline. These limitations can be reduced by incorporating openEHR local codes into the OMOP vocabulary and by submitting missing concepts to the appropriate terminology authorities.

A further difficulty arises from the interaction between TDCI and the heavy use of the observation table. Because OMOP stores many semantically distinct elements as observations or measurements, users must locate the correct codes among very large terminology sets before they can retrieve the desired information. A diagnosis archetype produces 10 observations, and the combination of diagnosis and problem qualifier produces 23. Each of these entries requires its own terminology code and its own position within the structure including a fact\_relationship to link it. As the number of tables increases, the number of codes that must be known and queried also increases. The openEHR ecosystem currently contains more than 800 archetypes \cite{CKM2025}, mapping these to OMOP faithfully would likely result in thousands of linked tables with diminishing utility.

This approach places greater reliance on terminology navigation and code selection compared to model-driven approaches. Even simple clinical constructs must be reconstructed by assembling multiple tables and identifying the correct terminology codes associated with each record. This places a substantial cognitive and technical burden on data scientists and magnifies the effect of any vocabulary gaps or local coding variations.

In openEHR, this information is contained within the archetype that defines the clinical concept. As a result, the data are organized in a consistent and interoperable structure and consolidated within a single archetype-defined structure, which can simplify retrieval of related elements once the model is known. The number of archetypes is currently 800 and each archetype presents a clear and explicit model of the concept, which avoids the proliferation of entries and codes that characterizes the TDCI-based approach in OMOP. MDCI is effective only because archetypes are developed and maintained through international governance, which ensures that the structures remain consistent across implementations and that the clinical meaning of each element is shared across sites. If archetypes are modeled locally this can lead to interoperability problems.

This difference is one of the main reasons why openEHR is adopted as a clinical data repository for persistence \cite{info:doi/10.2196/55779} and as national/regional record \cite{bajric2023building, piera202425, mclean2025new}. Its model driven concept identification, supported by international archetype governance, provides a stable and interoperable semantic foundation that is difficult to achieve in standards that rely on TDCI. Accordingly, the national Australian AUCDI initiative models its core data using archetypes and then derives corresponding FHIR profiles \cite{AUCDI}.

MDCI is not without limitations. Representing a new clinical concept requires a formally governed model, introducing modeling and governance overhead that TDCI avoids by relying on terminology codes. If models are defined locally rather than through international governance, semantic consistency across sites can no longer be guaranteed. Once formally governed, a clinical model can be reused indefinitely across sites, projects, and even derived into other standards such as FHIR profiles or OMOP, an investment that amortizes over time.

\subsection{Semantic ambiguity}

Some OMOP fields are semantically overloaded and combine several distinct clinical concepts into a single column. A prominent example is condition\_status\_concept\_id, which is used to represent admission diagnosis, primary diagnosis, postoperative diagnosis, and the cause of death. In clinical reality, these qualifiers are not mutually exclusive: an admission diagnosis may also be the primary diagnosis, and a primary diagnosis may also be recorded as a cause of death. Because OMOP provides only one status field, implementers are forced to select a single value for the main record and to represent all additional qualifiers through separate observation records linked via fact\_relationship (or similar mechanisms such as observation\_event\_id).

This design leads to loss of granularity and heterogeneous implementations. Different sites may choose different values to populate the main status field and move the remaining qualifiers into auxiliary tables, resulting in multiple structurally valid but semantically divergent representations of the same clinical scenario. Such variability may reduce interoperability and complicates cross-site and federated analyses, because it is no longer clear which qualifiers were retained in the core OMOP field and which were externalized into linked records.

\subsection{ETLs}

ETL processes introduce additional variability because several OMOP fields are not precisely defined from a clinical perspective. For example, onset date and clinically recognized time represent related but distinct concepts, yet their intended use is not clearly delineated. As a result, sites may interpret these fields differently and map identical source data in inconsistent ways, leading to divergent OMOP representations of the same clinical event.

Contextual information is also frequently lost during transformation. openEHR archetypes may distinguish multiple temporal aspects such as onset date, first appearance, and last appearance, whereas OMOP often provides only a single corresponding field. In the absence of explicit mapping rules, ETL processes may populate mandatory fields using default or context dates, further obscuring the original clinical meaning. Together, these ambiguities reflect the model's emphasis on analytic flatness over contextual representation  and reduce comparability across OMOP implementations.


\section{Discussion}
The OMOCL language was successfully extended with conceptMap logic to support the mapping of openEHR's internal terminologies. The Eos tool was updated to incorporate this logic. A new visit\_occurrence AQL logic was implemented to better adapt to different openEHR deployments. The international mapping library of OMOCL was also extended and now covers 198 archetype mappings. As a result, data contained in these archetypes is now processable in OMOP. This lowers the barrier to making clinical records stored in openEHR accessible for research.

The mapping effort represents the first comprehensive evaluation of openEHR archetypes against OMOP's model. A key finding is that archetypes often encapsulate more up-to-date and granular clinical knowledge than existing standard terminologies. The limited coverage of standardized codes (e.g. SNOMED CT, LOINC) within archetypes hampers automated transformation. To improve this situation, we made a list of the internal codings we mapped from openEHR to OMOP using terminologies and provided it to the openEHR community alongside the mappings \cite{OMOCL2025}. We hope that this increases the coverage of standardized terminology annotations in the international CKM.

Although the mapping library now covers 198 archetypes, more than 600 CKM archetypes remain unmapped and will need to be addressed as they reach stable status, either through provisional or complete mappings. The evaluation in this study is limited to structural and semantic representability and does not assess downstream analytical or clinical outcomes. A comprehensive downstream evaluation would require implementing representative cohort definitions, outcome phenotypes, and risk models on transformed datasets, ideally in multi-centric settings, and comparing their results against source-system queries as well as established OMOP-based implementations to assess consistency, reproducibility, and potential bias introduced during transformation. While performance was evaluated in our previous work, the approach has not yet been systematically benchmarked in large-scale or multi-center production environments. Finally, the reported rates of unmappable concepts cover the primary OMOP concept\_id fields and the conceptMap entries. They exclude additional contextual fields mapped via observation or fact\_relationship tables, which may further increase this proportion. A systematic quantification of how much reference model context OMOP can represent was beyond the scope of this study and remains a direction for future work.

\subsection{Structural and semantic limitations of the OMOP CDM}
Measured across the mapping library, only 12.8\% of the openEHR nodes have a direct field in the OMOP table their archetype maps to. The other 87.2\% need an auxiliary record, or OMOP cannot represent them at all. This limits the depiction of basic clinical concepts represented in openEHR archetypes across the library and not only in individual archetypes. The transformation of a single coherent openEHR clinical statement into OMOP therefore resulted in a graph of loosely coupled records, as illustrated by the detailed diagnosis archetype analysis. Semantic field overloading affected 35 of the 87 OMOP columns the library writes into (40.2\%), which combine conceptually distinct clinical notions into a single attribute. In most cases a single date column absorbs several clinically distinct timestamps, while the remaining columns force a choice between clinical notions. This design inevitably produces heterogeneous implementations across sites, as each ETL process must decide which qualifier to prioritize. Thus, even when OMOP schemas are identical, the resulting datasets may not be semantically aligned.

Contextual linking of related data is rarely supported in the OMOP tooling. If used, each additional contextual element requires a separate terminology concept, a separate OMOP record, and a linking relationship. For the diagnosis archetype alone, this results in ten linked observation records, rising to 23 when combined with the qualifier cluster. Given that the openEHR ecosystem contains more than 800 archetypes, faithful transformation would result in thousands of auxiliary records, creating a combinatorial explosion of codes and relationships that is difficult to govern and compromises query reliability. For these reasons, the fact\_relationship mechanism was implemented in a single mapping to demonstrate technical feasibility but was not systematically applied across the library.

These limitations could potentially affect the reliability of cohort construction, outcome classification, and risk stratification, thereby increasing the risk of subtle and potentially undetectable errors in downstream clinical analyses and study conclusions. Beyond the vocabulary gaps, which can be addressed by extending the vocabulary, the fundamental problems arise from OMOP's generic table structures and terminology-centric approach. OMOP has produced a mature open-source analytics ecosystem, including tools such as ATLAS \cite{Atlas} and Achilles \cite{hripcsak2015observational}, and underpins large international research networks such as EHDEN \cite{EHDEN}, enabling reproducible observational research at scale. Its generic, denormalised structure is a deliberate design, well suited for efficient large-scale, cross-site cohort queries, and we assess it against that purpose rather than against openEHR's modelling goals. While this makes OMOP a pragmatic approach that lowers the barrier to research collaboration, provides access to large datasets, and enables traceability, our mapping analysis shows that the same flattening means it may not fully preserve basic clinical concepts as represented in openEHR archetypes, even before considering more complex or specialized clinical models. This reflects an inherent trade-off of prioritising analytic flatness over granular clinical representation rather than a deficiency of the model. The mapping of openEHR archetypes, which are explicitly designed to represent clinical reality, reveals structural scalability challenges in the OMOP CDM and highlights gaps in its ability to represent several  clinically relevant concepts. Without changes to the underlying modeling strategy, particularly the need for explicit representation of clinical concepts and reduction of ambiguity introduced by generic tables and relationship links, these limitations may constrain OMOP's ability to represent highly granular clinical detail without further evolution of the model.

A path forward could involve closer alignment between the two ecosystems, drawing on the complementary strengths of each. The structural representation of clinical concepts could be informed by archetype-based modelling, which provides explicit, internationally governed clinical models that preserve context and reduce ambiguity. The terminology and analytics layer could build on OMOP's mature vocabulary system and its extensive open-source tooling for cohort analytics, data quality assessment, and federated research. These capabilities are currently underdeveloped in the openEHR ecosystem. Such an alignment reflects the principle that clinical concepts do not change based on their use: the same concept applies whether the data are used for care or for research. Realizing this convergence, whether through adoption of a curated subset of archetypes within OMOP, derivation of OMOP structures from archetypes, or a jointly governed intermediate model, would require sustained coordination between both communities. One potential challenge is that both archetypes and vocabularies evolve over time and are governed through formal versioning processes, introducing dependencies on both governance cycles that would need careful management.

Both openEHR and OMOP are well-established standards, and Eos together with OMOCL provides a practical migration path from openEHR to OMOP. openEHR is not only used for primary care documentation but is increasingly adopted for secondary use, including prospective clinical research. A convergence of both ecosystems would therefore benefit all stakeholders: institutions transforming their data to openEHR would progressively standardize their patient records in a vendor-neutral, semantically explicit format, while simultaneously enabling participation in large-scale secondary-use research networks based on OMOP. In this way, transformation to OMOP would become significantly easier, more reliable, and coherent between implementations when based on archetypes, making OMOP more suitable for modern clinical research while reducing the sources of bias and misinterpretation that currently affect clinical studies. However, such convergence would require coordinated governance, community alignment, and sustained adoption efforts across both ecosystems. Work in this direction is already underway through the OHDSI openEHR Working Group, which is actively addressing identified gaps including terminology coverage and structural alignment between the two standards. Ultimately, if successfully realized, this convergence could meaningfully reduce the translation overhead and information loss, shifting secondary-use data modeling toward a more faithful representation of clinical reality.


\section{Conclusion}
This work presents a new generation of the Eos engine and the OMOCL mapping language, substantially advancing interoperability between openEHR and OMOP. The addition of conceptMap logic, improved visit generation, and a significantly extended mapping library enable more accurate and comprehensive transformation of openEHR data and lower substantially the technical barriers for producing research-ready OMOP datasets. This enables more reproducible and scalable secondary use of clinical data.

Nevertheless, this work demonstrates that OMOP has structural and semantic limitations that go beyond technical mapping issues and may affect downstream analyses and therefore clinical study outcomes. Based on the mapping of 198 archetypes, including the diagnosis archetype which required 23 OMOP records to represent a single clinical statement, we measured structural constraints that limit semantic scalability and cause context loss. Coherent clinical concepts are fragmented across generic tables, clinically relevant qualifiers must be externalized through auxiliary records, and limited terminology coverage leads to persistent semantic gaps. These measured properties introduce representational ambiguity, increase the risk of misinterpretation in downstream analyses, and highlight scalability challenges that warrant attention as clinical research grows in complexity. Adopting an archetype-driven approach could provide a comprehensive, internationally governed clinical modeling framework that strengthens semantic clarity across implementations. Convergence between the openEHR and OMOP ecosystems offers a promising path toward improving semantic integrity in secondary-use data, but would require coordinated governance, community alignment, and sustained adoption efforts across both communities. If successfully realized, such convergence could shift secondary-use data modeling toward a semantically coherent and scalable representation of clinical reality.

\backmatter

\section*{List of abbreviations}
\begin{tabular}{ll}
\textbf{AQL} & Archetype Query Language \\
\textbf{CDM} & Common Data Model \\
\textbf{CKM} & Clinical Knowledge Manager \\
\textbf{EHR} & Electronic Health Record \\
\textbf{ETL} & Extract Transform Load \\
\textbf{FHIR} & Fast Healthcare Interoperability Resources \\
\textbf{LOINC} & Logical Observation Identifiers Names and Codes \\
\textbf{MDCI} & Model-Driven Concept Identification \\
\textbf{OMOCL} & OMOP Conversion Language \\
\textbf{OMOP} & Observational Medical Outcomes Partnership \\
\textbf{OMOP CDM} & Observational Medical Outcomes Partnership Common Data Model \\
\textbf{SNOMED-CT} & Systematized Nomenclature of Medicine Clinical Terms \\
\textbf{TDCI} & Terminology-Driven Concept Identification \\
\end{tabular}

\bibliography{sn-bibliography}

\section*{Declarations}

\subsection*{Ethics approval and consent to participate}
Not applicable.

\subsection*{Consent for publication}
Not applicable.

\subsection*{Availability of data and materials}
All source code and mapping resources supporting this study are openly available in public repositories.

\noindent\textbf{Eos} \cite{Eos2025}
\begin{itemize}
  \item Project name: Eos
  \item Project home page: https://github.com/SevKohler/Eos
  \item Operating system(s): Platform independent
  \item Programming language: Java
  \item Other requirements: JDK $\geq$ 17, Apache Maven $\geq$ 3.8.0, PostgreSQL 
  \item License: Apache-2.0
  \item Any restrictions to use by non-academics: None
\end{itemize}

\noindent\textbf{OMOCL} \cite{OMOCL2025}
\begin{itemize}
  \item Project name: OMOCL
  \item Project home page: https://github.com/SevKohler/OMOCL
  \item Operating system(s): Platform independent
  \item Programming language: OMOCL
  \item License: Apache-2.0
  \item Any restrictions to use by non-academics: None
\end{itemize}

\noindent\textbf{OMOCLanalytics} \cite{OMOCLanalytics2025}
\begin{itemize}
  \item Project name: OMOCLanalytics
  \item Project home page: https://github.com/SevKohler/OMOCLanalytics
  \item Operating system(s): Platform independent
  \item Programming language: Python
  \item License: Apache-2.0
  \item Any restrictions to use by non-academics: None
\end{itemize}

\subsection*{Competing interests}
S.K. and D.B. serve on the openEHR Specification Board (non-profit). D.B. is employed by VeraTech for Health S.L., a health informatics consultancy. The authors declare no further competing interests.

\subsection*{Funding}
This work was conducted within the HiGHmed consortium. The project is funded by the German Federal Ministry of Education and Research (BMBF, grant id: 01ZZ2302A, 01ZZ2302C).

\subsection*{Authors' contributions}
S.K. was the primary author of the manuscript, responsible for conceptualization, methodology, software and specification development, and validation. D.B. contributed to software and specification development, validation, and manuscript review. P.G.B. contributed to software development, validation, and manuscript review. F.M.E. contributed to methodology development, validation, and manuscript review. S.A. contributed to methodology review and manuscript review. R.E. and M.M. provided supervision and manuscript review. All authors read and approved the final manuscript.

\subsection*{Acknowledgements}
Not applicable.

\section{Appendix}

\begin{table}[ht]
\centering
    \caption{Archetypes without mapping and the reasons}
\begin{tabular}{|p{4.5cm}|p{4cm}|}
    \hline
    \textbf{Archetype} & \textbf{Reason} \\
    \hline
        CLUSTER.exam\_blastocyst.v1 & The specific exam is not available in the terminology. \\
    \hline
        CLUSTER.exam\_embryo.v1 & The specific exam is not available in the terminology.  \\
    \hline
        CLUSTER.exam\_oocyte.v1 & The specific exam is not available in the terminology.  \\
    \hline
        CLUSTER.exam\_zygote.v1 & The specific exam is not available in the terminology.  \\
    \hline
        CLUSTER.timing\_daily.v1 & Only part of another archetype. \\
    \hline
        CLUSTER.timing\_nondaily.v1 & Only part of another archetype. \\
    \hline
        EVALUATION.absence.v2 & Absence of information should not be represented in OMOP. \\
    \hline
        EVALUATION.social\_summary.v1 & Acts as a container for other archetypes (education, housing, etc.) which are mapped individually. \\
    \hline
        OBSERVATION.embryo\_assessment.v1 & Specific codes are not available in the terminology. \\
    \hline
        SECTION.adhoc.v1 & Archetype has no specific semantics. Ignore. \\
    \hline
\end{tabular}
\label{tab:archetypes_no_mapping}
\end{table}

\begin{table}[h!]
\caption{openEHR Problem/Diagnosis qualifiers represented in OMOP via OBSERVATION rows linked to CONDITION\_OCCURRENCE through FACT\_RELATIONSHIP}
\centering
\begin{tabular}{|p{4cm}|p{8cm}|}
\hline
\textbf{openEHR field} &
\textbf{OMOP table linked via FACT\_RELATIONSHIP} \\
\hline

Disorder category &
OBSERVATION \\
\hline

Current or past &
OBSERVATION \\
\hline

Active or inactive &
OBSERVATION \\
\hline

Level of control &
OBSERVATION \\
\hline

Progression &
OBSERVATION \\
\hline

Resolution phase &
OBSERVATION \\
\hline

Remission status &
OBSERVATION \\
\hline

Episodicity &
OBSERVATION \\
\hline

Reason for an ongoing episode &
OBSERVATION \\
\hline

Occurrence &
OBSERVATION \\
\hline

Course label &
OBSERVATION \\
\hline

Diagnostic category &
OBSERVATION \\
\hline

Admission diagnosis &
OBSERVATION \\
\hline

Comment &
NOTE \\
\hline

\end{tabular}
\label{tab:problem_qualifier_factrel}
\end{table}

\end{document}